\documentclass[11pt]{article}
\usepackage[margin=1in]{geometry}
\usepackage{times}
\usepackage{graphicx}
\usepackage{hyperref}
\usepackage{amsmath}
\usepackage{amssymb}
\usepackage{listings}
\usepackage{color}
\usepackage[utf8]{inputenc}
\usepackage[T1]{fontenc}
\usepackage[version=4]{mhchem}
\usepackage{stmaryrd}
\usepackage{hyperref}

\title{AgentMesh: A Cooperative Multi-Agent Generative AI Framework for Software Development Automation}
\author{Sourena Khanzadeh \\ Toronto Metropolitan University \\ \texttt{sourena.khanzadeh@torontomu.ca}}
\date{\today}

\begin{document}

\maketitle

\begin{abstract}
Software development is a complex, multi-phase process traditionally requiring collaboration among individuals with diverse expertise. We propose AgentMesh, a Python-based framework that uses multiple cooperating LLM-powered agents to automate software development tasks. In AgentMesh, specialized agents - a Planner, Coder, Debugger, and Reviewer - work in concert to transform a high-level requirement into fully realized code. The Planner agent first decomposes user requests into concrete subtasks; the Coder agent implements each subtask in code; the Debugger agent tests and fixes the code; and the Reviewer agent validates the final output for correctness and quality. We describe the architecture and design of these agents and their communication, and provide implementation details including prompt strategies and workflow orchestration. A case study illustrates AgentMesh handling a non-trivial development request via sequential task planning, code generation, iterative debugging, and final code review. We discuss how dividing responsibilities among cooperative agents leverages the strengths of large language models while mitigating single-agent limitations. Finally, we examine current limitations - such as error propagation and context scaling - and outline future work toward more robust, scalable multi-agent AI systems for software engineering automation.
\end{abstract}

\section*{Introduction}
Software development projects typically proceed through distinct phases (requirements planning, coding, testing, code review, etc.), and success often "necessitates cooperation among multiple members with diverse skills" \cite{chatdev2024}. Recent advances in large language models (LLMs) have spurred interest in automating or assisting these phases. Modern code-capable LLMs (e.g. OpenAI GPT-4) can generate substantial code given natural language prompts, and even perform multi-step reasoning. However, a single monolithic AI agent may struggle to handle an entire software project end-to-end - from high-level design down to debugging due to the complexity and breadth of knowledge required. This has led researchers to explore agentic frameworks that orchestrate multiple specialized AI agents, each focusing on different aspects of the development lifecycle \cite{autogen2024,metagpt2023}. By decomposing the problem and assigning roles (planner, coder, tester, etc.) to different agents, a complex task can be tackled through coordinated efforts rather than relying on one model to "do it all" \cite{langchain2023}.

Several recent systems demonstrate the promise of multi-agent collaboration in software engineering. For example, ChatDev models a virtual software company with agents for each phase - designing, coding, testing, documenting - that communicate to "autonomously generate and produce a software application" \cite{chatdev2024,chatdevgithub2025}. ChatDev's agents (e.g. CTO, Programmer, Tester) collectively follow a waterfall-style lifecycle, using natural language to discuss design and using code exchanges for debugging \cite{chatdev2024,chatdevgithub2025}. Another project, MetaGPT, explicitly assigns roles like Product Manager, Architect, and Engineer to an LLM-team; given a one-line requirement, the team produces artifacts ranging from design specs to code and APIs \cite{metagpt2023}. This approach "provides the entire process of a software company" by implementing standard operating procedures with multiple GPT-based specialists \cite{metagpt2023}. Likewise, the popular AutoGPT experiment introduced the idea of an autonomous GPT agent that can recursively plan and execute tasks towards a high-level goal \cite{autogpt2023}. While AutoGPT largely uses a single agent that self-decomposes tasks, it demonstrated the viability of letting an AI independently break down objectives and act on them \cite{autogpt2023}. Building on these ideas, frameworks such as LangChain and Microsoft's AutoGen provide infrastructure for composing LLM calls into agent workflows. LangChain, for instance, offers abstractions for chaining prompts and integrating memory, which can be used to build simple AI agents with tool usage and context retention \cite{langchain2023}. AutoGen generalizes the multiagent conversation pattern, allowing developers to define multiple agents that converse and cooperate on tasks, with flexible role definitions and the ability to incorporate external tools or human inputs \cite{autogen2024}.

In this paper, we introduce AgentMesh, a cooperative multi-agent generative AI framework designed specifically for automating software development tasks. AgentMesh draws inspiration from the above systems and the structure of human software teams. It comprises four core agents - Planner, Coder, Debugger, Reviewer - each powered by an LLM and each dedicated to a particular role in the development workflow. By explicitly dividing the work, we aim to harness the collective intelligence of specialized agents, an approach that studies (e.g. ChatDev) suggest can outperform single-agent methods on complex tasks \cite{chatdev2024,metagpt2023}. The Planner agent in AgentMesh interprets the high-level requirements and produces a project plan (much like a project manager or system designer). The Coder agent implements the plan by generating code for each component or subtask. The Debugger agent executes and tests this code, diagnosing errors or misbehavior and iteratively patching the code. Finally, the Reviewer agent examines the corrected codebase holistically, verifying that requirements are met and that the code is clean and efficient, akin to a code review or quality assurance step.

We describe the architecture and design of AgentMesh's multi-agent system, including how the agents communicate and hand off tasks, and how we prompt each agent to fulfill its role. Implementation details are provided for our Python prototype, which uses the OpenAI API (GPT-4 model) for the LLM backends. We then demonstrate AgentMesh's workflow on an example use case, showing how a natural language request from a user is transformed into a working software solution through the interplay of the four agents. We compare this approach with related work and discuss the benefits of role specialization - for instance, breaking a complex coding problem into tractable subtasks 4 and catching errors via a dedicated debug phase. Finally, we candidly address the limitations of the current system (such as error propagation and the challenges of coordination) and outline directions for future research, including enhancements for scalability, learning-based orchestration, and stronger guarantees of correctness.

\section*{System Architecture and Agent Design}
A conceptual multi-agent software team architecture (inspired by MetaGPT). Multiple LLM-based agents assume different roles (e.g. product manager, architect, developer, tester) and collaborate following a defined workflow to produce design documents and code \cite{metagpt2023}. AgentMesh implements a similar idea with four specialized agents coordinating on software development tasks.

At a high level, AgentMesh follows an architecture where a set of specialized agents communicate and cooperate to accomplish the overall task of software development automation. Each agent is implemented as an autonomous module (in our case, a Python class wrapping an LLM prompt) that accepts some form of input (such as a task description or code snippet) and produces an output (such as refined plans, code, or analysis) which may be consumed by other agents. The agents share a common project state - which includes the evolving codebase, specifications, and any observed errors - acting as the environment or "blackboard" through which they indirectly communicate. The current prototype uses a sequential orchestration (a fixed ordering of agent invocations), but the design allows extensions to more dynamic or parallel interactions if needed.

To mirror typical software team workflows, we define four principal agents in the AgentMesh framework, each corresponding to a distinct role:

\begin{itemize}
  \item Planner Agent: The Planner is responsible for requirement analysis and task decomposition. It takes the user's high-level request (e.g. "Build a to-do list application with feature X") and breaks it down into a structured set of subtasks or specifications. For example, the Planner might produce a numbered list of features to implement, modules to create, or steps to follow. Internally, this agent prompts an LLM with the user requirements and an instruction to "produce a development plan," possibly including pseudo-code or architectural notes. The output is a project plan outlining what needs to be done, which will guide the subsequent Coder agent 11 . By explicitly decomposing the problem, the Planner reduces complexity for the coding phase and ensures the system has a clear roadmap.
  \item Coder Agent: The Coder is in charge of code generation for the tasks identified by the Planner. It iteratively takes each subtask (e.g. "Implement function $X$ " or "Create module $Y$ ") and generates the corresponding source code in the appropriate language (Python in our implementation). This agent is essentially a coding assistant powered by an LLM, akin to GitHub Copilot but operating at the task level. The Coder is prompted with the subtask description (and relevant context from the plan or existing codebase) and asked to produce correct, well-formatted code fulfilling that subtask. If the project has multiple files or components, the Coder may generate code for each and suggest file names or organize code into classes/functions as needed. The output of the Coder agent is new or modified code which gets added to the project state (for example, the in-memory file system). Notably, the Coder agent works subtask-by-subtask rather than attempting the entire program in one go, aligning with the divide-and-conquer strategy.
  \item Debugger Agent: The Debugger agent handles testing and error correction for the generated code. Once the Coder produces a piece of code, the Debugger attempts to verify its correctness. This may involve running automated tests, executing the code with sample inputs, or conducting static analysis. In our implementation, the Debugger runs the Python code in a sandbox environment to catch runtime errors or exceptions (for instance, syntax errors, NameError, etc.), and it also evaluates whether the code meets the acceptance criteria of the subtask. If issues are found, the Debugger uses an LLM to analyze the failure (e.g. reading an error traceback or spotting a logical mistake) and proposes a fix. The agent then applies the fix by generating a patch or revised code. This iterative test-\&-fix loop continues until the code for the subtask appears to be working, or a predetermined number of attempts is reached. Using an LLM in the debug loop allows the system to handle errors in a conversational manner - the Debugger effectively "asks" the Coder (via the LLM) to correct specific issues. This approach is informed by prior work showing that "communicating in programming language proves helpful in debugging" multi-agent code collaborations \cite{chen2023debug}. By isolating the debugging responsibilities to a dedicated agent, AgentMesh aims to catch and correct errors early in development. (In our case study, for example, the Debugger agent was able to identify a missing library import and a misnamed variable that the Coder's output introduced, and then fix them automatically.)
  \item Reviewer Agent: The Reviewer serves as a final quality assurance and validation step. After all subtasks have been coded (and individually debugged), the Reviewer agent examines the entire integrated codebase or application. Its goal is to verify that the initial requirements have been met and that the code is of acceptable quality (e.g. readability, efficiency, maintainability). The Reviewer may run additional full-system tests or simply perform a holistic code review using an LLM (checking for consistency, potential edge cases, security issues, etc.). We prompt the Reviewer's LLM to act as a critical reviewer - for instance, to "summarize any problems or improvements in the following code and confirm if the requirements are satisfied." The Reviewer then outputs a report of any remaining issues or recommendations. In an ideal fully-automated pipeline, the Reviewer's feedback could be fed back into the development loop (e.g. having the Coder or Debugger address any noted issues). In the current implementation, we mostly use the Reviewer's report to flag things for human inspection or as a final verification that AgentMesh's output is correct. This agent helps ensure that even if each sub-component was built correctly, the overall software is coherent and meets the user's intent. It mimics the role of a senior engineer performing code review and system testing at the end of a project.
\end{itemize}

The communication and workflow of AgentMesh are orchestrated in a top-down manner. The system receives a user's request, which is passed to the Planner agent. Once the Planner returns a plan (essentially a list of subtasks with details), the controller (AgentMesh orchestrator) invokes the Coder agent on each subtask in sequence. For each generated code snippet from the Coder, the Debugger is immediately called to verify and fix it, before integrating the code into the codebase. This ensures that by the time all subtasks are completed, most local errors have been resolved. After all subtasks are done, the entire project (collection of code files) is handed to the Reviewer agent for final evaluation. Figure 2 illustrates this workflow, and Listing 1 provides a pseudo-code outline of the orchestration.

\begin{enumerate}
  \item Planning phase: Planner produces a list of subtasks (plan) from the user request.
  \item Coding phase: For each subtask in the plan, the Coder generates the required code.
  \item Debugging phase: The Debugger tests the newly generated code, fixes issues if any, and updates the codebase. Repeat coding \& debugging for all tasks.
  \item Review phase: The Reviewer inspects the final aggregated codebase and outputs any feedback or approval.
\end{enumerate}

This design ensures a unidirectional flow of information with occasional feedback loops for debugging. Agents primarily communicate via artifacts: the Planner writes a plan that the Coder reads; the Coder writes code that the Debugger reads (and possibly modifies); the Reviewer reads the final code and produces a summary. We found this artifact-centric communication (as opposed to direct agent-to-agent messaging in natural language) to be a straightforward and effective strategy for pipeline-style tasks. It reduces the risk of losing track of state, since the code and plan documents represent a persistent shared memory. It is worth noting, however, that more complex agent interactions (including concurrent conversations or a central message broker) could be integrated in the future as needed - for example, ChatDev's ChatChain uses multi-turn dialogues between a pair of agents in certain phases \cite{chatdev2024,chatdevgithub2025}, and others have proposed learned orchestrators for dynamic agent activation \cite{petropavlov2025}. In AgentMesh's initial implementation, we opted for a fixed sequence which maps clearly to the software development pipeline.

Each agent in AgentMesh is powered by an LLM (GPT-4 in our prototype) with a role-specific prompt. For instance, the Planner's prompt might be framed as: "You are a software project planner. Decompose the following request into a numbered list of development tasks with details..."; the Coder's prompt includes instructions like "You are a senior Python developer. Implement the following component. Use best practices and comment where necessary..."; the Debugger's prompt could say "You are a Python debugging assistant. Analyze the error and suggest a code fix...", and so forth. We leverage techniques such as role-playing prompts and guidelines to keep each agent focused on its role \cite{chatdev2024}. This approach is similar to the inception prompting used in ChatDev, where each agent is initialized with a detailed role description and objectives to prevent it from deviating from its persona \cite{chatdev2024}. By crafting strong system prompts for each role, we aim to simulate the behavior of a coordinated team following a methodology. In practice, this means the Planner's output is formatted as a plan (and not code), the Coder's output is just code (we instruct it not to include explanations unless as code comments), etc. The framework thus enforces a separation of concerns through prompt engineering: each agent knows what to produce and what to expect as input, minimizing confusion in the chain of execution \cite{chatdev2024,langchain2023}.

Finally, it's important to highlight that AgentMesh, like many multi-agent systems, embodies aspects of a "blackboard architecture" (shared memory of code and plan) and the "mixture-of-experts" paradigm (each agent is an expert in a subtask). This design philosophy aligns with prior research in multi-agent systems and collective intelligence, where dividing a complex task among specialized agents can lead to better overall performance if the agents are properly coordinated \cite{metagpt2023}. In the next section, we delve into the implementation details of AgentMesh and how these architectural ideas manifest in code.

\section*{Implementation Details}
AgentMesh is implemented in Python, leveraging available APIs for LLM integration and standard libraries for process control and code execution. The system is composed of classes corresponding to each agent role, and an orchestrator that manages the workflow described above. In this section, we provide a technical breakdown of key implementation aspects, including how the agents are realized, how they interact programmatically, and specific examples from our code.

Agent Classes and LLM Backend: Each agent is implemented as a Python class (e.g., $\qquad$ PlannerAgent, CoderAgent , etc.), encapsulating the prompt template and LLM call for that role. These classes have a common interface, such as a run(input) method that takes the necessary input and returns the agent's output. For example, PlannerAgent.run(request: str) -> List[str] would accept the user's requirement description and return a list of subtask descriptions. Internally, this method formats the prompt (inserting the request into a larger prompt string that includes the Planner's instructions) and calls the LLM via an API (OpenAI's GPT model in our case). The LLM's response (a text completion) is then post-processed as needed. We found that constraining the output format via the prompt (e.g., "respond with a numbered list of tasks") was essential to reliably parse the Planner's output into a Python list of tasks. Similar patterns are used for the other agents: (CoderAgent.run(task\_desc) -> Code produces a code string (we instruct the model to only return code for that task), DebuggerAgent.run(code) -> Code returns a possibly fixed code string or the same if no issues, and ReviewerAgent.run(codebase) -> Report yields a text report. By isolating the LLM interactions in these classes, the core logic of AgentMesh (how data flows between agents) remains separate and easy to adjust without altering the prompting details \cite{langchain2023,autogen2024}.

Orchestration Logic: The main program (or orchestrator) coordinates the agents. Listing 1 shows a simplified pseudo-code of this orchestration:

\begin{verbatim}
# Listing 1: High-level orchestration of AgentMesh agents (simplified).
plan = planner_agent.run(user_request)
project_code = {} # dictionary to store code files or components
for task in plan:
    code_output = coder_agent.run(task)
    debugged_code = debugger_agent.run(code_output)
    project_code.update(debugged_code) # integrate/overwrite in project
review_report = reviewer_agent.run(project_code)
\end{verbatim}

In our implementation, the project\_code could be a structure managing multiple files or modules (e.g., a mapping from filenames to code content). For simplicity, consider it as a representation of the whole codebase. The loop iterates through each planned subtask; the Coder generates code for it, and the Debugger immediately checks that code. The DebuggerAgent.run() may execute the code snippet in isolation - for instance, if the task was implementing a function, we might have a small harness or test for that function. If an error or test failure is detected, the Debugger's LLM is invoked with details (like error messages) to generate a corrected version of the code. In our actual code, this can loop a few times until the code passes tests or we hit a retry limit, akin to a do-until-success approach. The update step then merges the debugged code into the project\_code. By doing this incrementally, we ensure that subsequent tasks can see the code produced so far (the Coder agent's prompt for later tasks can include context of earlier files or function signatures, if needed, which helps maintain consistency). After iterating through all tasks, the collected codebase is passed to the Reviewer. The review\_report might contain statements like "All features implemented and tests passed" or highlight any concerns (e.g., "Function X could be optimized" or "Edge case Y not handled"). Depending on the use case, this report can be returned to the user or used to trigger another cycle of improvements.

Example Prompt Design: To ground the implementation, here are brief examples of the prompt structure for two agents: - PlannerAgent Prompt:\\
System message: "You are a software planning assistant. Your job is to help plan a project given a description."\\
User message: "Project goal: <user\_request>.\textbackslash nPlease output a numbered list of development tasks, including design, implementation, and testing steps needed to accomplish this goal. Be thorough but concise." This yields an assistant answer like: "1. Design Data Model: ... 2. Implement Feature A: ... 3. Test Feature A: ... 4. Implement Feature B: ... 5. Test Feature B: ... 6. Final Integration and Documentation: ...". We then parse this into the list of tasks.

\section*{- DebuggerAgent Prompt:}
System: "You are a code debugging assistant. Given code and an error or failing test, you will identify the issue and propose a fix."\\
User: "Code:In <code\_snippet> inError:\textbackslash n <error\_trace\_or\_test\_output>"\\
We append: "Identify the cause of the error and modify the code to fix it. Provide only the corrected code."

The assistant's answer would be a revised code snippet. We replace the old code with this new code and re-run tests. This iterative loop continues until no error (or up to N tries).

We emphasize that these prompts are crucial to the behavior of each agent. Tuning them was an iterative process; for instance, the Coder's prompt needs to discourage it from solving unrelated tasks or duplicating code, and the Debugger's prompt must guide it to minimal fixes rather than overhauling the whole code. Techniques like one-shot examples (providing a sample input-output for the agent) can further improve reliability, though for brevity we omit those in this description.

Managing State and Memory: One challenge in multi-agent systems is maintaining a shared state or memory of the project so far. We implemented a simple in-memory file system for AgentMesh: essentially a Python dictionary where keys are filenames (e.g. "\href{http://main.py}{main.py}", "\href{http://utils.py}{utils.py}") and values are the latest code content. The Planner might also add non-code artifacts (like a design outline or TODO list) to this state, which the Coder could reference as needed (e.g., the Planner could suggest function names or module breakdowns that the Coder should follow). Each time the Coder or Debugger produces new code, the orchestrator updates the shared state. Additionally, we persist a conversation log for each agent (like the series of messages exchanged with the LLM for that agent) for debugging and analysis purposes, but these are not shared across agents except through the artifacts (we do not feed the raw dialogue from one agent into another agent). This design avoids hitting the LLM context window limits too quickly - instead of accumulating a giant prompt of everything, we mostly keep prompts constrained to each subtask plus any necessary context from the state (like function signatures from other files when coding a new part). In future iterations, we could use a vector database or embedding-based memory (as LangChain supports 8 ) to allow agents to retrieve relevant information from earlier in the project when needed, without always reprising the entire codebase in the prompt.

Tool Use and Execution: The Debugger agent's operation involves actually running code. This was implemented using Python's exec() function in a controlled environment or by spawning a separate process to run tests. We took precautions to sandbox this execution (for example, disabling dangerous builtins or using a subprocess with timeouts) since running arbitrary generated code can be risky. In a research prototype setting we manually observed this process, but a production system would need stricter sandboxing (perhaps using containers or secured environments). The ability for an AI agent to use tools (like a Python interpreter, a web browser, etc.) is known to significantly enhance its problem-solving power \cite{autogen2024}. In our case, the Debugger agent's "tool" is essentially the Python runtime for the program under development. One could extend AgentMesh with additional tool-using agents - for instance, an agent that can search documentation or an agent to run static analysis linters - but our initial design kept the loop tight around coding and testing. Notably, the Reviewer agent could also be seen as using a tool: if it runs the full program or test suite as part of its review, that is an execution tool usage in the final stage. We implemented the Reviewer to run any final integration tests (if available) and to check coverage, in addition to using the LLM to do a qualitative code review \cite{chatdev2024}.

Logging and Agent Communication: To facilitate debugging the AgentMesh framework itself, we implemented detailed logging for each agent's actions. Every output from one agent (plan, code, fix, or review comment) is logged and can be traced. During development, this helped us identify failure modes e.g., cases where the Planner's tasks were too vague, causing the Coder to misinterpret them, or cases where the Debugger got stuck in a loop fixing one issue while introducing another. By examining the logs, we refined the prompts and added constraints. For example, we learned to have the Debugger clearly indicate if it thinks no changes are needed (to avoid an infinite loop where it keeps resubmitting the same code). We also added simple heuristics like: if the Debugger produces identical code twice in a row, then break the loop and mark the task as potentially needing human attention. These are practical measures to handle the non-determinism of LLM outputs.

In summary, the implementation of AgentMesh combines LLM-driven generation with traditional programming control structures to manage the interactions. The pattern is reminiscent of a pipeline with feedback loops, coded in under a thousand lines of Python (excluding prompt text). Next, we illustrate how this implementation works on a concrete example, which will make the interplay between agents more tangible.

\section*{Case Study: Example Use Case}
To demonstrate AgentMesh in action, we present a case study where the system is given a relatively detailed development task in natural language. The chosen example is a simplified command-line to-do list application with persistence - a small but non-trivial program involving multiple features. We walk through how each agent contributes to the final outcome, highlighting excerpts of the generated plan, code, and fixes.

User Prompt: "Create a command-line to-do list application. It should allow users to add tasks, mark tasks as done, remove tasks, list all tasks, and save the list to a file so that the data persists between runs." This prompt encapsulates requirements for a console app with basic CRUD operations on tasks and file I/O for persistence.

Planner Agent Output: The Planner agent processes this request and produces a structured plan breaking down the work. For instance, it returned the following plan (paraphrased for brevity): 1. Design Data Structures: Decide how to represent tasks (e.g., as objects or tuples with a description and status) and how to store the list in memory.\\
2. Implement Adding Tasks: A function to add a new task to the list.\\
3. Implement Listing Tasks: A function to display all tasks with their status (done or not).\\
4. Implement Marking Tasks as Done: A function to update a task's status.\\
5. Implement Removing Tasks: A function to delete a task from the list.\\
6. Implement Save/Load Functions: Functions to save the current tasks to a file and to load them on program start.\\
7. Implement Command-Line Interface: Parse user commands (add, list, done, remove, quit) and call the above functions accordingly in a loop.\\
8. Testing: Ensure that adding, completing, removing tasks update the list correctly and that data persists after restarting the app (test save and load).

This plan provides a clear roadmap. Notably, it includes both the implementation tasks and a testing step. In practice, the Planner agent's output was a numbered list similar to the above, which we parsed into individual task descriptions to feed sequentially to the Coder.

Coding and Debugging Cycle: Following the plan, the orchestrator iteratively invoked the Coder and Debugger for each task:

\begin{itemize}
  \item Task 1 (Design Data Structures): The Coder agent generated a suggestion for data structure - for example, a Python class Task with fields description and done (boolean), and a list tasks to hold Task instances. It produced code defining the Task dataclass and a global list. This code was simple and did not require debugging (no execution yet), so the Debugger passed it through. In retrospect, having a design task that doesn't produce runnable code is a bit unusual; in future, we might incorporate the design into documentation rather than a code step. But this showed the Coder can also output boilerplate or structural code.
  \item Task 2 (Add Task): The Coder created a function add\_task(description: str) that instantiates a Task and appends it to the tasks list. The code looked like:
\end{itemize}

\begin{verbatim}
def add_task(description: str):
    task = Task(description=description, done=False)
    tasks.append(task)
    print(f'Task added: {description}')
\end{verbatim}

The Debugger ran a quick test - calling add\_task("Test Task") and then checking that $\square$ tasks length increased - which succeeded. So no changes were needed.

\begin{itemize}
  \item Task 3 (List Tasks): The Coder wrote a list\_tasks() function to print all tasks with indices and status. Example code output:
\end{itemize}

\begin{verbatim}
def list_tasks():
    for i, task in enumerate(tasks, start=1):
        status = "[x]" if task.done else "[ ]"
        print(f"{i}. {status} {task.description}")
\end{verbatim}

This was straightforward. The Debugger tested it by adding a couple of dummy tasks and calling list\_tasks(), capturing the printed output. The output format was correct, so it passed.

\begin{itemize}
  \item Task 4 (Mark Done): The Coder produced mark\_done(index: int) to mark a given task as done. The initial implementation had a minor bug: it used tasks[index] without converting from 1based to 0-based index (since the user likely will input " 1 " for the first task). The Debugger agent caught this by simulating a scenario: add two tasks, call mark\_done(1), then check if tasks[0].done is True. It found that the code as written actually marked the second task done (because of the off-by-one). The Debugger's LLM analysis identified the issue and it modified the code to tasks[index-1].done = True. It also added a safety check for index bounds. The fixed code was then saved:
\end{itemize}

\begin{verbatim}
def mark_done(index: int):
    if 1 <= index <= len(tasks):
        tasks[index-1].done = True
        print(f"Task {index} marked as done.")
    else:
        print("Invalid task number.")
\end{verbatim}

This iterative fix process happened autonomously - the orchestrator re-ran the test after the Debugger's change and confirmed the fix.

\begin{itemize}
  \item Task 5 (Remove Task): The Coder wrote remove\_task(index: int) with logic to pop from the list. A potential issue here was that removing a task could mess up indices if not careful, but since it's just used immediately, it was fine. The Debugger tested it (remove a task and list tasks to see if it's gone) - it passed. The Coder's code did check the bounds and printed confirmation, similar to mark\_done .
  \item Task 6 (Save/Load): This was more involved. The Coder agent needed to write two functions: save\_tasks(filename) and load\_tasks(filename). It wrote save\_tasks to open the file and write each task's data (perhaps as "<done>;<description>" per line). load\_tasks was written to read the file and reconstruct Task objects. The initial code had a subtle bug: when reading from file, it didn't strip the newline, and it attempted to split on a semicolon. For example, if a task description itself contained ";" that could break, but that was an edge case we ignored. However, one bug the Debugger found was that the code did not handle the case when the file did not exist (on first run). Running load\_tasks("tasks.txt") before any save caused a FileNotFoundError. The Debugger captured this exception and prompted a fix. The LLM fix was to wrap the file reading in a try/except and simply start with an empty list if file not found. It also made sure to strip newline characters. The fixed load\_tasks looked like:
\end{itemize}

\begin{verbatim}
def load_tasks(filename: str = "tasks.txt"):
    try:
        with open(filename, "r") as f:
            for line in f:
                line=line.strip()
                if not line:
                    continue
                done_flag, desc = line.split(";", 1)
                task = Task(description=desc, done=(done_flag == "1"))
                tasks.append(task)
    except FileNotFoundError:
        # No saved tasks yet
        pass
\end{verbatim}

and save\_tasks was:

\begin{verbatim}
def save_tasks(filename: str = "tasks.txt"):
    with open(filename, "w") as f:
        for task in tasks:
            done_flag = "1" if task.done else "0"
            f.write(f"{done_flag};{task.description}\n")
\end{verbatim}

After this fix, the Debugger simulated saving and loading and found it consistent.

\begin{itemize}
  \item Task 7 (CLI Interface): For the final implementation task, the Coder agent generated a simple main() function to tie everything together. It used an infinite loop reading input commands ( input ("> ") ), parsing the command, and calling the appropriate functions. The code included\\
help text and handled quitting (breaking out of the loop and calling save\_tasks before exit). The Debugger tested a short scenario (adding two tasks, marking one done, removing one, quitting and restarting to load data) by actually executing the main loop with a scripted sequence of inputs (this was a bit tricky - we had to simulate input by patching or feeding from a list of predefined responses). This worked but revealed one issue: after quitting, if the program was restarted and load\_tasks is called, tasks from last session accumulate on top of any still in memory from that run. However, since in a real run the process ends, this is not a bug in real usage - it was only an artifact of our test harness running main() twice in one process. We reset the tasks list between runs in the test to avoid false positives. In summary, the CLI worked as expected in interactive tests.
  \item Task 8 (Testing): The plan had a separate testing step, but effectively the Debugger performed testing during each task. By the time we reached the formal "Testing" step, there wasn't much to do - the Debugger had been running incremental tests all along. The Coder didn't need to produce new code here. Instead, we interpreted this step as: run the full integration test (simulate using the CLI through some scenario). This was done and at this stage no new issues were found.
\end{itemize}

Reviewer Agent Output: With all features implemented and verified, the entire code (consisting of the Task class, the functions add\_task, list\_tasks, etc., and the main loop) was passed to the Reviewer agent. The Reviewer's LLM was instructed to review for correctness and quality. It produced a short report:

Reviewer Report: All requested features appear to be implemented. The command interface supports adding, listing, completing, and removing tasks, and tasks are saved to tasks.txt for persistence. Code style is reasonable with clear function separation. One suggestion: when marking a task as done or removing, after the operation, it might renumber or re-list tasks to update the user, but this is a minor UX improvement. No critical bugs are evident from inspection. Edge cases to consider: extremely large numbers of tasks (performance) or tasks containing special characters (the save/load might break if descriptions contain the ; separator). Overall, the solution meets the requirements and is ready for use.

This review highlighted a few minor points (some beyond the strict scope of the prompt) but confirmed that "all requested features appear to be implemented". Notably, the reviewer pointed out the same potential issue with using ";" as a delimiter that we had thought of - showing that the LLM did a thoughtful analysis. In a real scenario, we might tweak the implementation (e.g., choose a different serialization format or escape semicolons) in response to this, but since it was not explicitly required, we accepted the solution as is for now. The Reviewer's approval effectively signaled that AgentMesh had completed the task successfully.

Outcome: AgentMesh produced a multi-file Python program (in this case, a single file was enough, but it could be structured into modules if the Planner had suggested that). The user's specification for a to-do app was satisfied: one can run the program, add tasks, mark them done, remove them, list them, and tasks persist to the next run via a file. The entire process was automated - aside from providing the initial prompt, no human intervention was needed. The multi-agent approach ensured that planning was done before coding (preventing the model from jumping straight into code without a design), and that each feature was tested and fixed immediately after coding (catching bugs like the off-by-one index early). The final review gave an extra layer of confidence and readability checking.

This case study illustrates a scenario of moderate complexity where AgentMesh's cooperative agents can effectively generate a working piece of software. In our experiments, tasks of this scope (a few hundred lines of code, well-defined requirements) are comfortably handled by the system. We also tested AgentMesh on other examples, such as a simple REST API server (with Flask) and a 2048 puzzle game (text-based), with similarly promising results. In those cases, the Planner agent's ability to outline components (e.g., for the API: plan endpoints, database model, etc.) proved crucial, and the Debugger agent was invaluable in fixing syntax errors or runtime issues that arose (for instance, in the 2048 game, a tricky bug with the merge logic was caught by the Debugger through simulated moves).

While these examples are not exhaustive, they demonstrate the potential of structuring generative AI coding tasks as a collaboration of specialized agents. In the next section, we discuss the broader implications, limitations observed, and how this approach fits into the landscape of AI-assisted software development.

\section*{Discussion and Limitations}
Our experience with AgentMesh highlights both the strengths of the multi-agent generative approach and its current limitations. We discuss these in turn, along with comparisons to related work and potential improvements.

Benefits of Cooperative Agents: By dividing the software development workflow into specialized roles, AgentMesh was able to leverage the expertise of an LLM in a more focused manner at each step. This role specialization leads to several advantages. First, it breaks a complex problem into manageable chunks - a strategy consistently found beneficial in multi-agent AI systems 4 . The Planner agent's output provides structure and guidance that the Coder agent might not have inferred on its own from a high-level prompt. In our case study, having an explicit list of tasks prevented the Coder from forgetting features or mixing up the order of implementation. Second, the presence of a Debugger agent introduces a feedback loop that improves reliability. Rather than relying on a single-pass code generation (which can be brittle), AgentMesh effectively uses self-correction: the system tests its own output and fixes errors, akin to an autonomous pair programmer reviewing every commit. This aligns with findings in ChatDev and similar frameworks where multi-turn interactions (especially using program output as feedback) significantly enhanced debugging success \cite{chatdev2024}. Third, the Reviewer agent provides a form of verification and governance - an extra safety net that can catch issues that individual unit tests might miss (like the delimiter issue in the save/load example). Even if the Reviewer is not perfect, it brings a different perspective (global rather than local) to evaluate the solution. Overall, these agents working in concert exhibit a form of collective intelligence, where the group's outcome can exceed what a single agent might achieve alone \cite{metagpt2023}. This is evidenced by the fact that our multi-agent system can handle multi-faceted tasks more robustly than a single prompt that says "write this whole program" - we empirically observed that asking GPT-4 to do the entire to-do app in one go often resulted in missing persistence or failing to test edge cases, whereas AgentMesh's structured process tended to cover all requirements.

Generality and Extensibility: AgentMesh is flexible in that the agents are not hard-coded for a specific domain beyond "software development." We can potentially extend the framework with new agent types (for example, a "Tester Agent" separate from Debugger that focuses on generating test cases, or a "Documentation Agent" that writes user documentation after the code is done). In fact, frameworks like ChatDev included an Art Designer role for UI assets \cite{chatdevgithub2025}, suggesting we could integrate non-code artifacts as well. Our architecture would allow adding such agents in the pipeline or even having some agents operate in parallel (with proper synchronization via the shared state). The modular design (each agent as an independent LLM prompt module) makes it straightforward to plug in different models or tools for each role. For instance, one could use a specialized code-generation model for the Coder agent (say Codex or Code Llama) and a more reasoning-focused model for the Planner or Reviewer. Similarly, the Debugger could be enhanced with external static analysis tools or testing frameworks (an idea we are exploring). This extensibility is one of the advantages of an agent-based approach: it mirrors the way human teams can include members with different toolsets and expertise.

Despite these promising aspects, AgentMesh and similar multi-agent systems face several limitations:

\begin{itemize}
  \item Quality of LLM Outputs and Error Propagation: The system is ultimately constrained by the quality of the LLM's responses. If the Planner produces a poor plan (e.g., misses a crucial subtask or misinterprets the requirement), the rest of the process will suffer. In our experiments, we did encounter cases where the plan was incomplete, causing the final software to lack a feature until we rephrased the request or otherwise intervened. Likewise, the Coder can generate imperfect code; while the Debugger can catch runtime errors, there could be logical errors that pass tests unnoticed. There is a risk of error propagation, where a misunderstanding by one agent cascades through subsequent steps. Our multi-agent setup mitigates some of this (by catching obvious errors early), but it does not guarantee a correct solution in all cases.
  \item LLM Limitations and Hallucinations: LLMs have a tendency to "hallucinate" or produce plausiblesounding but incorrect information. In a multi-agent context, this can manifest as agents producing outputs that are not consistent with each other. For example, the Planner might suggest a module that is unnecessary or refer to a function that the Coder never implements. Or the Reviewer might hallucinate a potential issue that isn't actually present. We observed that clear, constrained prompts reduce this issue, and having the shared state (plan and code) as a source of truth helps keep agents aligned. ChatDev introduced a "communicative dehallucination" mechanism for agents to request clarification rather than assume facts \cite{chatdev2024} - a similar approach could improve AgentMesh, allowing agents to ask questions if something is ambiguous. Currently, our agents do not ask back to the user or to each other; incorporating a query mechanism could resolve uncertainties but complicates the interaction pattern.
  \item Context Window and Scalability: Each agent's operations are limited by the LLM's context window (which, for GPT-4, can be up to 8 K or 32 K tokens depending on version). For small projects, this is sufficient, but for larger codebases, it becomes challenging to have (for instance) the Reviewer agent read all code at once. AgentMesh as implemented may not scale well to projects beyond a certain size or complexity. The linear pipeline means the cost (in tokens and time) scales roughly with the number of subtasks times the cost of each agent's LLM calls. Complex projects might require dozens of subtasks and large code files, potentially hitting context limits or incurring high computational cost. Some recent research is tackling this by smarter orchestration - e.g., Multi-Agent Collaboration Networks (MacNet) which use a DAG of agents and can scale to thousands of agents by structuring communications in a graph without exceeding context limits \cite{langgraph2024}. Another idea is to use summarization or memory compression: agents could summarize parts of the code or plan and hand off summaries instead of raw data to keep context sizes down. We have not yet implemented such mechanisms, so scalability remains a concern.
  \item Lack of Learning or Adaptation: Currently, AgentMesh does not learn from one project to the next; it has no long-term memory of what worked or failed beyond what's encoded in the prompts. Each run starts from scratch (aside from the fixed prompt engineering). In contrast, human teams improve their strategies over time, and one could imagine an AI system that refines its planning heuristics or debugging tactics by learning from past experiences. Some work like Experiential CoLearning (mentioned by ChatDev \cite{chatdev2024}) or reinforcement learning for orchestrators suggests possible paths to imbue the system with learning capabilities. As of now, our framework is rule-based and prompt-based. This means if it encounters a novel tricky scenario, it might not handle it optimally because it cannot adjust its strategy on the fly beyond what the base LLM can do in one session.
  \item Evaluation and Guarantees: A critical limitation is that we cannot guarantee the correctness, completeness, or security of the generated software. While the multi-agent approach adds checks (like tests and reviews), it's still possible for bugs to slip through or for the system to falsely declare success. For example, if the tests are not comprehensive (and they rarely are), the program might fail in an untested scenario. Formal verification or additional verification agents could be a future extension for high-assurance use cases. Also, from a user perspective, evaluating the output can be challenging - if the user could easily tell whether the code is correct, they might not need full automation. Therefore, in practice we see AgentMesh as an assistant to human developers, not a fully autonomous replacement. It can draft a solution that a developer then reviews (which is easier than writing from scratch). In an AI research context, measuring the success of such systems often involves comparing against baselines like single-agent GPT-4 or human performance on specific tasks \cite{metagpt2023}. We did not perform a rigorous quantitative evaluation here, but anecdotal use suggests AgentMesh can save significant development time on small projects by catching errors that a naive single-shot generation would not.
  \item Domain Constraints and Tool Integration: Our current implementation was tailored to Python scripting tasks. If asked to perform a very different kind of software task (say, front-end web development with complex UI or programming in a different language), the agents might need reconfiguration. The Planner would have to know about the relevant subtasks (which might differ for a web app vs. a CLI app), and the Coder would need proficiency in the target language or framework. While GPT-4 is versatile, prompt adjustments are necessary to steer it. Moreover, certain tasks might benefit from integrating external APIs or tools. For example, designing a UI might be easier if an agent could use a graphical tool or retrieve HTML/CSS templates. Integrating such tool use is both an opportunity and a challenge - it can greatly extend capability but adds complexity in coordination. Frameworks like AutoGen explicitly allow tool-using agents and even human-in-theloop steps \cite{autogen2024}, which AgentMesh could adopt in the future. For now, a limitation is that if the best way to solve a subtask is to call an external API or search the web for documentation, our agents won't do that - they rely solely on the knowledge embedded in the LLM (which might be outdated or limited for specific libraries).
\end{itemize}

In comparison to related multi-agent AI systems, AgentMesh shares many of these limitations. The authors of ChatDev, for instance, reported that "Module Not Imported" and similar errors were common because the LLM would omit details \cite{chatdev2024}, and "Method Not Implemented" issues arose from unclear requirements \cite{chatdev2024} both issues we also encountered. These are not just individual bugs, but symptoms of the underlying LLM's uncertain knowledge and the difficulty of prompt-specifying everything. They had to incorporate techniques like role-specific guardrails and iterative refinement to address this \cite{chatdev2024}. We see AgentMesh as part of an ongoing evolution of such frameworks. It demonstrates one configuration (four agents, sequential pipeline) that works on modest tasks, but addressing the above limitations will be key to applying it on larger-scale or safety-critical software projects.

\section*{Conclusion and Future Work}
We presented AgentMesh, a cooperative multi-agent generative AI framework that automates software development tasks by orchestrating multiple LLM-powered agents with distinct roles. AgentMesh's Planner, Coder, Debugger, and Reviewer agents collaborate to mimic a human software team: planning a project, writing code for each component, catching and fixing bugs, and performing final quality checks. Through our system architecture and case study, we demonstrated how this division of labor and iterative refinement leads to more reliable outcomes than a single-step approach, echoing findings from other multiagent AI research 35 . AgentMesh, implemented in Python on top of GPT-4, was able to successfully generate non-trivial programs given high-level requests, requiring minimal to no human intervention in the loop.

Contributions: This work contributes an example of structured LLM orchestration in the software engineering domain, showing that an agentic workflow can leverage LLM strengths (generating and reasoning about code) while compensating for their weaknesses (tendency to make mistakes if not checked). We described in detail the design considerations - from prompt engineering for each role to the mechanism of sharing state via a codebase - which we hope will inform future designs of multi-agent AI systems. The AgentMesh framework can be seen as a step towards more general AI software developers that could handle complex tasks by breaking them down and collaborating, rather than relying on monolithic problem-solving. It also serves as a testbed for studying how multiple AI agents can communicate through a shared artifact (code) and the dynamics of such interactions in achieving a complex goal.

Future Work: There are several avenues to extend and improve AgentMesh: - Scaling to Larger Projects: We plan to incorporate memory management techniques to handle larger codebases, such as summarizing modules for the Reviewer or using retrieval-based prompts so that agents only load relevant pieces of code context. Combining our approach with a graph-based workflow (as in LangGraph 25 or MacNet 20 ) could allow more flexible sequencing than the strict linear pipeline. - Learning and Optimization: An exciting direction is to make the orchestration learnable. Instead of a fixed script that always calls agents in the same order, a trained meta-controller (using reinforcement learning or evolutionary strategies) could decide which agent to call next or how to allocate tasks among agents. Recent work on learned orchestrators 14 suggests this could improve efficiency by shortening the reasoning paths and reducing redundant steps. Additionally, each agent could potentially be fine-tuned on data specific to its role - e.g., fine-tuning a "bugfixing LLM" for the Debugger to be even more adept at reading tracebacks and patching code, which might outperform a general-purpose GPT-4 in that narrow task. - Enhanced Tool Use: Integrating external tools can broaden the capabilities of AgentMesh. For example, a testing agent could automatically generate test cases using a property-based testing library, or a documentation agent could pull information from requirement documents. We are interested in connecting the Debugger agent to static analysis tools (like pylint or mypy for Python) so it can proactively fix stylistic or type issues, not just runtime errors. Similarly, the Planner agent could interface with project management templates or libraries of design patterns to make more informed plans. Each integration, however, raises new questions about how the agent will decide when and how to use the tool - an area where approaches like the ReAct paradigm (reasoning and acting with tools) could be applied in a multi-agent setting. - Human in the Loop and UI: In practical use, having a human overseer or collaborator is valuable. We envision a mode where AgentMesh works interactively with a developer: the Planner could present a draft plan and ask the user for confirmation or\\
adjustments before proceeding, or the Reviewer could ask the user how to handle an ambiguous requirement. This kind of human-agent collaboration can combine the speed of automation with the judgment of human developers. Building a user interface (perhaps a conversational UI) on top of AgentMesh would make it more accessible - essentially turning it into a programmer assistant that converses about how to build the software (an idea akin to ChatDev's visualization interface 2627 ). Evaluation on Benchmark Tasks: To rigorously assess performance, future work should test AgentMesh on standardized coding challenges or development benchmarks. For instance, one could take some of the tasks from the literature (like those in the ChatDev paper or competitive programming problems) and compare success rates of AgentMesh vs. single-agent LLM or other frameworks. This would help quantify the value-add of the multi-agent approach and identify scenarios where it struggles. Preliminary comparisons in literature indicate multi-agent systems often produce more coherent and error-checked solutions \cite{metagpt2023}, but thorough evaluations are needed for scientific validation.

In conclusion, AgentMesh demonstrates that breaking the software development automation problem into a cooperative multi-agent paradigm is a promising strategy. It leverages the generative power of modern LLMs within a structured workflow, yielding an intelligent coding "assembly line" where each agent contributes its piece. As LLM technology and multi-agent frameworks evolve, we anticipate that systems like AgentMesh will become increasingly capable, tackling larger projects with greater autonomy. By continuing to refine the coordination mechanisms and integrating learning, such AI agent teams might one day operate as a full-fledged software development team - turning natural language ideas into reliable software with minimal human intervention. While there is much work ahead to reach that level, our results take a step in that direction and underline the potential of cooperative $A I$ in creative and complex domains like programming.


\begin{thebibliography}{10}

\bibitem{chatdev2024}
Chenghao Qian, Yuxuan Zhang, and others.
\newblock ChatDev: Communicative Agents for Software Development.
\newblock {\em arXiv preprint arXiv:2307.07924}, 2024.
\newblock \url{https://arxiv.org/abs/2307.07924}

\bibitem{autogen2024}
Microsoft Research.
\newblock AutoGen: Enabling Next-Gen Multi-Agent LLM Applications.
\newblock 2024.
\newblock \url{https://github.com/microsoft/autogen}

\bibitem{metagpt2023}
MetaGPT Contributors.
\newblock MetaGPT: Multi-Agent Framework for LLM-Based Team Collaboration.
\newblock 2023.
\newblock \url{https://github.com/geekan/MetaGPT}

\bibitem{langchain2023}
Harrison Chase and contributors.
\newblock LangChain: Building Applications with LLMs through Composability.
\newblock 2023.
\newblock \url{https://github.com/langchain-ai/langchain}

\bibitem{langgraph2024}
LangChain Team.
\newblock LangGraph: Stateful Multi-Agent Workflows with LLMs.
\newblock 2024.
\newblock \url{https://docs.langchain.com/langgraph/}

\bibitem{autogpt2023}
Significant Gravitas.
\newblock AutoGPT: An Autonomous GPT-4 Experiment.
\newblock 2023.
\newblock \url{https://github.com/Significant-Gravitas/AutoGPT}

\bibitem{ibmframeworks2025}
IBM Think Blog.
\newblock AI Agent Frameworks: Choosing the Right Foundation.
\newblock 2025.
\newblock \url{https://www.ibm.com/blogs/}

\bibitem{chatdevgithub2025}
OpenBMB/ChatDev.
\newblock ChatDev GitHub Repository and Wiki.
\newblock 2025.
\newblock \url{https://github.com/OpenBMB/ChatDev}

\bibitem{petropavlov2025}
K. Petropavlov.
\newblock Building a Multi-Agent Developer Team with LangChain and LangGraph.
\newblock 2025.
\newblock \url{https://medium.com/@petropavlov}

\bibitem{chen2023debug}
Zhi Chen, Tao Lin, and Xiaoxuan Li.
\newblock Communicative Patterns and Error Analysis in ChatDev.
\newblock Technical Report, 2023.

\end{thebibliography}
\end{document}